# Simpson's Paradox and the implications for medical trials

Norman Fenton, Martin Neil and Anthony Constantinou

31 August 2015

Abstract

This paper describes Simpson's paradox, and explains its serious implications for randomised control trials. In particular, we show that for any number of variables we can simulate the result of a controlled trial which uniformly points to one conclusion (such as 'drug is effective') for every possible combination of the variable states, but when a previously unobserved confounding variable is included every possible combination of the variables state points to the opposite conclusion ('drug is not effective'). In other words no matter how many variables are considered, and no matter how 'conclusive' the result, one cannot conclude the result is truly 'valid' since there is theoretically an unobserved confounding variable that could completely reverse the result.

## 1. Introducing Simpson's paradox

"Simpson's paradox" is a commonly occurring statistical phenomenon first formally recorded in 1951 [6] and has been addressed in [7] and [5]. For reasons that we will make clear in this paper, it is not really a paradox even though it can often appear to look like one.

A classic and simple example of Simpson's paradox is the scenario in which Jane's scores are worse than Fred's every year but she still comes with better overall average. How is this possible?

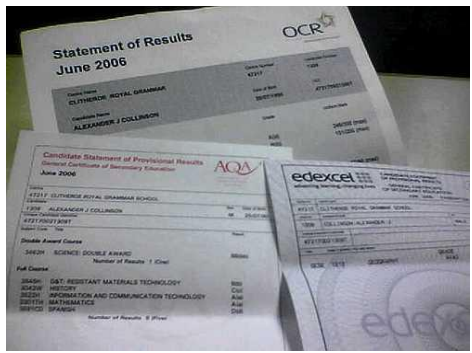

Fred and Jane study on the same course spread over two years. To complete the course they have to complete 10 modules. At the end, their average annual results are as shown in Table 1.

Table 1 Course results

|  | *Year 1 average* | *Year 2 average* | *Overall Average ?* |
|---|---|---|---|
| Fred | 50 | 70 | 60 |
| Jane | 40 | 62 | 51 |

So how is it possible that Jane got the prize for the student with the best grade?

It is possible because the overall average figure is an average of the year averages rather than an average over all 10 modules. We cannot work out the average for the 10 modules unless we know how many modules each student takes in each year.

In fact:

- Fred took 7 modules in Year 1 and 3 modules in Year 2
- Jane took 2 modules in Year 1 and 8 modules in Year 2.

Assuming each module is marked out of 100, we can use this information to compute the total scores as shown in Table 2

**Table 2 Revised score information**

|  | *Year 1 total* | *Year 2 total* | *Overall total* | *Real Overall average* |
|---|---|---|---|---|
| Fred | 350 (7 x 50) | 210 (3 x 70) | 560 | 56.0 |
| Jane | 80 (2 x 40) | 496 (8 x 62) | 576 | 57.6 |

So clearly Jane really did do better overall than Fred.

This is an example of *Simpson's paradox*. It seems like a paradox – Fred's average marks are consistently higher than Jane's average marks but Jane's overall average is higher. But it is not really a paradox. It is simply a mistake to assume that you can take an average of averages without (in this case) taking account of the number of modules that make up each average.

Look at it the following way and it all becomes clear: in the year when Fred did the bulk of his modules he averaged 50; in the year when Jane did the bulk of her modules she averaged 62. When you look at it that way it is not such a surprise that Jane did better overall.

There have been many well-known cases where Simpson's paradox has clouded rational judgement and decision making. One of the most famous occurred at Berkeley University [1], which was (wrongly) accused of sex discrimination on the grounds that its admissions process was biased against women. Overall the data revealed a higher rate of admissions for men, but no such bias was evident for any individual department. The overall bias was explained by the fact that more women than men applied to the more popular departments (i.e. those with a high rejection rate).

## 2. When Simpson's paradox becomes more worrying

Simpson's paradox is worrying because the most frequent instances in practice occur in medical studies. Consider the following example (based on one due to Pearl [4]).

A new drug is being tested on a group of 800 people (400 men and 400 women) with a particular disease. We wish to establish whether there is a link between taking the drug and recovery from the disease. As is standard with drug trials half of the people (randomly selected) are given the drug and the other half are given a placebo. The results in Table 3 show that, of the 400 given the drug, 200 (i.e. 50%) recover from the disease; this compares favourably with just 160 out of the 400 (i.e. 40%) given the placebo who recover.

**Table 3 Drug trial results**

| Drug taken    | No  | Yes |
|---------------|-----|-----|
| *Recovered*   |     |     |
| No            | 240 | 200 |
| Yes           | 160 | 200 |
| *Recovery rate* | 40% | 50% |

So clearly we can conclude that the drug has a positive effect. Or can we? A more detailed look at the data results in *exactly the opposite* conclusion. Specifically, Table 4 shows the results when broken down into male and female subjects.

**Table 4 Drug trial results with sex of patient included**

| Sex | Female | | Male | |
|---|---|---|---|---|
| Drug taken | No | Yes | No | Yes |
| *Recovered* | | | | |
| No | 210 | 80 | 30 | 120 |
| Yes | 90 | 20 | 70 | 180 |
| *Recovery rate* | 30% | 20% | 70% | 60% |

Focusing first on the men we find that 70% (70 out of 100) taking the placebo recover, but only 60% (180 out of 300) taking the drug recover. *So, for men, the recovery rate is better without the drug*.

With the women we find that 30% (90 out of 300) taking the placebo recover, but only 20% (20 out of 100) taking the drug recover. *So, for women, the recovery rate is also better without the drug*. **In every subcategory the drug is worse than the placebo**.

The process of drilling down into the data this way (in this case by looking at men and women separately) is called *stratification*. Simpson's paradox is simply the observation that, on the same data, stratified versions of the data can produce the opposite result to non-stratified versions. Often, there is a *causal* explanation. In this case men are much more likely to recover naturally from this disease than women. Although an equal number of subjects overall were given the drug as were given the placebo, and although there were an equal number of men and women overall in the trial, the drug was *not* equally distributed between men and women. More men than women were given the drug. Because of the men's higher natural recovery rate, overall more people in the trial recovered when given the drug than when given the placebo.

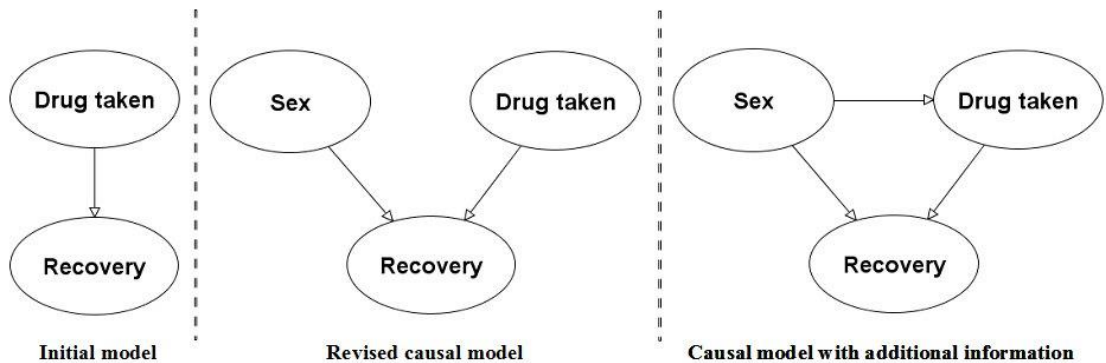

**Figure 1 Explaining Simpson's paradox using a causal model**

The difference between the types of data analysis is captured graphically in Figure 1. In the initial model we only have information about whether or not the drug is taken to help us determine whether a subject recovers or not. The revised causal model tells us that we need information about the subject's sex, in addition to whether they take the drug, to help us better determine whether the subject recovers. The final model introduces the further dependence, which is relevant *for this particular case study* namely that sex influences drug taken because men are **much more likely in this study** to have been given the drug than women.

### 3. Can we avoid Simpson's paradox?

The answer to this question is yes, but only if we are certain that we know every possible variable that can impact the outcome variable. If we are not certain – and in general we simply cannot be – then Simpson's paradox is theoretically unavoidable.

First let us see how we can avoid the paradox in the previous drup example. Another way of looking at the paradox in that example is that, because there are an unequal number of men and women in the study, the variable 'sex' **confounds** the recovery rate. In general, a variable can confound the results of a study if it is related (non-independent) to both the dependent variable ('recovery' in this case) and at least one of the other (independent) variables in the study ('drug taken' in this case). The only way to be sure of avoiding Simpson's paradox in our conclusions is if there are an equal number of subjects for each state of the confounding variable (in this case there are two 'states' of 'sex' namely male and female) for each state of the other dependent variable ('drug taken').

So, in the example study it is not sufficient to simply divide the subjects into two equal size 'control groups' (400 taking the drug and 400 taking the placebo) **even if the total number of males and females are equal**. We actually need four equal size control groups corresponding to each state combination of the variables, i.e.:

- 200 subjects who fit the classification ('drug', 'male')
- 200 subjects who fit the classification ('drug', 'female')
- 200 subjects who fit the classification ('placebo', 'male')
- 200 subjects who fit the classification ('placebo', 'female')

So, let us suppose that in a new study for some different drug we ensure that our 800 subjects are assigned into equal size control groups and that the results are as shown in Table 5.

**Table 5 New drug trial results with sex of patient included**

| Drug taken | No | Yes |
|---|---|---|
| Recovered | | |
| No | 192 | 148 |
| Yes | 208 | 252 |
| Recovery rate | 52% | 63% |
| Overall result: Favours drug | | |

| Sex | Female | | Male | |
|---|---|---|---|---|
| Drug taken | No | Yes | No | Yes |
| Recovered | | | | |
| No | 108 | 92 | 84 | 56 |
| Yes | 92 | 108 | 116 | 144 |
| Recovery rate | 46% | 54% | 58% | 72% |
| In each subcategory: Favours drug | | | | |

Note the following:

- All 4 control groups have 200 subjects.
- The overall recovery rate is 63% with the drug compared to 52% with the placebo
- The recovery rate among men is 72% with the drug compared to 58% with the placebo

- The recovery rate among women is 54% with the drug compared to 46% with the placebo

So, in contrast to the previous example, the drug is more effective overall and more effective in every sub-category. So surely we can recommend the drug and cannot possibly fall foul of Simpon's Paradox in this case?

Unfortunately, it turns out that we really can fall foul of the paradox - as soon as we realise there may be another confounding variable that is not explicit in the data. Consider the variable *age*, and for simplicity let us classify people with respect to this variable into just two categories "<40" and "40+". Even if we are lucky enough to have exactly 400 of the subjects 'in each category we may have a problem. Look at the results in Table 6 when we further stratify the data of Table 5 by age.

**Table 6 New drug trial results with sex and age of patient included**

| Age | 40+ | | | | < 40 | | | |
|---|---|---|---|---|---|---|---|---|
| Sex | Female | | Male | | Female | | Male | |
| Drug taken | No | Yes | No | Yes | No | Yes | No | Yes |
| Recovered | | | | | | | | |
| No | 96 | 28 | 80 | 24 | 12 | 64 | 4 | 32 |
| Yes | 64 | 12 | 80 | 16 | 28 | 96 | 36 | 128 |
| Recovery rate | 40% | 30% | 50% | 40% | 70% | 60% | 90% | 80% |

Since it is the *same* data obviously none of the previous results are changed, i.e. a higher proportion of people overall recover with the drug than the placebo, a higher proportion of men overall recover with the drug than the placebo, and a higher proportion of women overall recover with the drug than the placebo A. However, we can now see:

- *The proportion of young men who recover with the drug is 80% compared to 90% with the placebo*
- *The proportion of old men who recover with the drug is 40% compared to 50% with the placebo*
- *The proportion of young women who recover with the drug is 60% compared to 70% with the placebo*
- *The proportion of old women who recover with the drug is 30% compared to 40% with the placebo*

So, **in every single subcategory** the drug is actually **less effective** than the placebo. This is despite the fact that in every single super-category the exact opposite is true. How is this possible? Because, just as in the earlier example, the size of the control groups at the lowest level of stratification are not equal; more young people were given the drug than old people (320 against 80). And young people are generally more likely to recover naturally than old.

The only way to be sure of avoiding the paradox in this case would be to ensure we had 8 equal size control groups:

- 100 subjects who fit the classification ('drug', 'male', 'young')
- 100 subjects who fit the classification ('drug', 'female', 'young')
- 100 subjects who fit the classification ('placebo', 'male', 'young')
- 100 subjects who fit the classification ('placebo', 'female', 'young')
- 100 subjects who fit the classification ('drug', 'male', 'old')
- 100 subjects who fit the classification ('drug', 'female', 'old')

- 100 subjects who fit the classification ('placebo', 'male', 'old')
- 100 subjects who fit the classification ('placebo', 'female', 'old')

## 4. Generating arbitrary large examples of Simpsons paradox using a BN

In fact we can use Bayesian Networks (BNs) both to very easily construct arbitrarily large examples of Simpsons paradox and also to show how to counter the resulting problem of requiring unfeasibly large trials.

We create a model with *n* 'independent' parents X1, X2, …, Xn plus one parent D ('Drug taken') and one child 'Recovered' as shown in Figure 2 (assume all nodes are Boolean with states 'true' and 'false').

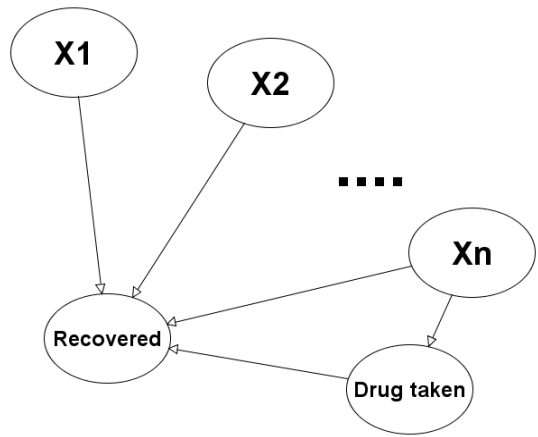

**Figure 2 Generalised Simpsons paradox**

What this model will simulate is the situation in which Xn is the confounding variable.

In the Node Probability Table for 'Recovered' define, for each state combination of X1, X2, …Xn

- P(Recovered = True | X1, X2, …Xn-1, Xn=False, D= False) = p1, where p1 is just above 0.5 (e.g. 0.52) for each possible state combination of X1, X2, …Xn-1
- P(Recovered = True | X1, X2, …Xn-1, Xn= True, D= False) = p2, where p2 is close to 1 (e.g. 0.9) for each possible state combination of X1, X2, …Xn-1
- P(Recovered = True | X1, X2, …Xn-1, Xn= False, D= True) = p3, where p3 is just below 0.5 (e.g. 0.48) for each possible state combination of X1, X2, …Xn-1
- P(Recovered = True | X1, X2, …Xn-1, Xn= True, D= True) = p4, where p4< p2 but is still close to 1 (e.g. 0.8) for each possible state combination of X1, X2, …Xn-1

For example, when n=3 we have the NPT:

| Drug taken | False | | | | | | | | True | | | | | | | |
|---|---|---|---|---|---|---|---|---|---|---|---|---|---|---|---|---|
| Xn | False | | | | True | | | | False | | | | True | | | |
| X2 | False | | True | | False | | True | | False | | True | | False | | True | |
| X1 | False | True | False | True | False | True | False | True | False | True | False | True | False | True | False | True |
| False | 0.48 | 0.48 | 0.48 | 0.48 | 0.1 | 0.1 | 0.1 | 0.1 | 0.52 | 0.52 | 0.52 | 0.52 | 0.2 | 0.2 | 0.2 | 0.2 |
| True | 0.52 | 0.52 | 0.52 | 0.52 | 0.9 | 0.9 | 0.9 | 0.9 | 0.48 | 0.48 | 0.48 | 0.48 | 0.8 | 0.8 | 0.8 | 0.8 |

Now define the NPT of D (drug taken) to be:

- P(D= True | Xn = True) = p
- P(D= True | Xn = False) = q

where p is close to 1 and q is close to 0. For example:

| Xn | False | True |
|---|---|---|
| False | 0.999 | 0.001 |
| True | 0.001 | 0.999 |

In other words, when Xn is True it is almost certain that the drug is taken and when Xn is False it is almost certain the drug is not taken. Now let us compare what happens when Xn *is* and *is not observed* (using the above example values).

Case 1: Xn *is observed*

By the definition of the NPT of Recovered, for every possible state combination X1, X2, …Xn we will have probability of recovery higher when D is False compared with when D is True, i.e. taking the drug is worse in every possible category.

Specifically, for every possible combination of the states X1, X2, …Xn-1 when Xn is True:

P(Recovered = yes |X1, X2, …Xn-1, Xn=True, D=True)  = 0.8

P (Recovered = yes |X1, X2, …Xn-1, Xn=True, D=False)  = 0.9

And for every possible state combination of X1, X2, …Xn-1 when Xn is False, we will have

P(Recovered = yes |X1, X2, …Xn-1, Xn=False, D=True)  = 0.48

P (Recovered = yes |X1, X2, …Xn-1, Xn=False, D=False)  = 0.52

Case 2: Xn *is not observed*

Let us assume that the prior probability of Xn is 50:50. Then, when we run the BN model, for every possible state combination of X1, X2, …Xn-1 we will have

P(Recovered = yes | X1, X2, …Xn-1, D=True)  is just less than 0.8

P (Recovered = yes | X1, X2, …Xn-1, D=False)  is just above 0.52

i.e. for every state combination, taking the drug has a higher probability of success than not taking it.

So, we have provided a method which, for an arbitrary number of variables, can ensure that when a single extra confounding variable is added every single result gets inverted. Because of this possibility we dispute the assertion in [5] that Simpson's paradox is 'resolved'.

## 5. The implications for randomized control trials (RCTs)

If we assume that all variables are Boolean then each time we find a new confounding factor we are obliged to double the number of trial participants. Considering factors such as various pre-existing medical conditions, lifestyle factors and ethnicity, it is clear that in general there could be hundreds of such variables. Even if we focused on the 20 'most important' we would need $2^{20}$ equal size control groups. That is over a million. To achieve statistically significant conclusions we would need at least 50 in each group – totalling more than the entire population of the UK. And if (as is likely for many attributes) we need a more fine-grained classification, such as *age* in years:

[0-10, 11-18, 19-20, 21-30, 31-40, 41-50, 51-60, 61-70, 71-80, >80]

then, even making the change for that single variable, means we require over 260 million subjects (close the population of the USA). Moreover, many of the control groups will contain 'rare' combinations of variable states meaning there is unlikely to be sufficient living people in the whole of the world to populate the study.

It is clear that relying on the availability of so many subject and data alone to learn meaningful relationships may be impractical. No matter how 'big' your trial is you may always need something even bigger. Even if we found the necessary 260 million subjects and associated data in the above example, if there is just one more confounding variable that was 'hidden' then the results of our data analysis may be perfectly wrong.

The only way to resolve these kinds of problems (and the related ones described in [3]) is to incorporate expert judgment. Indeed the very 'act' of identifying additional confounding variables is itself dependent on expert judgment (typically informed by related observational data). Figure 1 gives an indication of how causal (Bayesian network) models can help minimize the data requirements; the final causal model enables up to explicitly capture the fact that – in the data – men were more likely to be given the drug than women. As explained in [2] this enables us to make conclusions that avoid Simpsons paradox without being forced to get additional data with equal size control groups. Further methods for avoiding the paradox in practice (without falling into the big data black hole) are provided in works such as [4] and [7].

## Acknowledgements

This work was conducted as part of the European Research Council project ERC-2013-AdG339182-BAYES_KNOWLEDGE. See https://www.eecs.qmul.ac.uk/~norman/projects/B_Knowledge.html